# Generation of double-charged optical vortices on the basis of electrooptic Kerr effect

Yurij Vasylkiv, Ihor Skab and Rostyslav Vlokh *

Institute of Physical Optics, 23 Dragomanov St., 79005 Lviv, Ukraine
* Corresponding author: vlokh@ifo.lviv.ua

**Abstract**

We show that double-charged optical vortices can be generated with the help of Kerr electrooptic effect in either single crystals or isotropic media, including gaseous and liquid ones. We analyze possibilities for the vortex generation via the Kerr effect for different point group of symmetry and formulate the appropriate conditions. We prove that the crystals, textures and the isotropic media most suitable for the generation of double-charged optical vortices should belong to the symmetry groups 622, 6mm, 6/mmm, 6, 6/m, $\infty/m$, $\infty$, $\infty 2$, $\infty mm$, $\infty/mmm$, $\infty/\infty/mmm$, and $\infty/\infty 2$.

## 1. INTRODUCTION

In the last decade generation of singular beams with dislocations of their wave fronts [1] has become an important branch of optical studies. This is caused by wide applications of beams bearing optical vortices with a nonzero orbital angular momentum (OAM) [2]. The vortex beams can be successfully applied in a number of novel branches, e. g. information processing [3], quantum cryptography [4, 5], and quantum teleportation [6]. Furthermore, they can also be used for manipulating microparticles [7].

Scalar field singularities containing the optical vortices can be quite simply generated using spiral plates [8], computer-synthesized holograms [9], or optical wedges [10]. However, these methods for generating helical modes are 'passive', i.e. they do not permit operating spatial positions of the vortex beams, whereas the efficiency of induction of the OAM is predetermined by the structures used for the vortex beam generation. From the other side, polarization field singularities generated mainly in anisotropic media reveal a number of additional advantages, if compared with the singularities of scalar fields. For example, radially polarized beams can be used for achieving resolutions below the diffraction limit [11, 12]. Moreover, a phenomenon of spin-to-orbit angular momentum (SAM-to-OAM) conversion can be efficiency used in quantum coding-decoding, using qubits or even qudits of information [13, 14].

If one generates the vortex beams using so-called q-plates, representing in fact liquid crystal cells with structural defects in their centers, the efficiency of the SAM-to-OAM conversion can be tuned by adjusting temperature or electric field [15–17]. Recently we have suggested and verified experimentally several methods for the SAM-to-OAM conversion based on piezooptic effect induced in single crystals by torsion or bending [18–20]. Besides, electrooptic Pockels effect induced in single crystals by a conically shaped electric field has been found to be a convenient method for SAM-to-OAM converting and operating the efficiency of the vortex beam generation by an electric field applied [21, 22]. Let us remind that the Pockels effect can be utilized while controlling electrooptically the entanglement of SAM and OAM [23], thus being applicable for quantum teleportation.

On the other hand, the Pockels effect can exist only in acentric crystals, since a third-rank polar tensor describing the effect is equal to zero in any centrosymmetric media, including all isotropic ones. Moreover, as shown in the study [22], doughnut modes can be generated via the Pockels effect only in the crystals containing three-fold or six-fold inversion axes among their symmetry operations. Let us also remind that all of the methods of vortex beam generation that rely on single crystals can induce single-charged optical vortices only [18–22]. Furthermore, due to symmetry limitations only the bending method of vortex beam generation can be realized for isotropic materials, though even this method cannot be used if the working media are liquid or gaseous. However, an electrooptic Kerr effect can exist in these materials. The aim of the present work is to study the optical vortex generation via the Kerr effect in the material media of different symmetries.

## 2. PHENOMENOLOGICAL DESCRIPTION OF THE KERR EFFECT INDUCED BY CONICALLY SHAPED ELECTRIC FIELD

It is well known that the electrooptic effect manifests itself in changes of optical impermeability coefficients $B_i$ (or refractive indices $n$, since $B_{ij} = (1/n)^2_{ij}$) under the action of electric field $E$ (see, e.g., [24]). This effect is described by a tensorial relation

$$B_{ij} = r_{ijk} E_k + R_{ijkl} E_k E_l, \tag{1}$$

where $r_{ijk}$ and $R_{ijkl}$ are the tensors of Pockels and Kerr coefficients, respectively. Further we will use the matrix notation, i.e. the Kerr coefficients will be represented as $R_{lm} = R_{ijkl}$ for $ij \leftrightarrow l = 1,...6;\ kl \leftrightarrow m = 1, 2, 3$ and $R_{lm} = 2R_{ijkl}$ for $ij \leftrightarrow l = 1,...6;\ kl \leftrightarrow m = 4, 5, 6$.

As shown in the work [22], a conically shaped electric field (see Fig. 1) results in the following coordinate dependences of the electric field components:

$$E_1 = \frac{U}{d}\frac{Z}{X^2+Y^2+Z^2}X = mX, \quad E_2 = \frac{U}{d}\frac{Z}{X^2+Y^2+Z^2}Y = mY, \quad E_3 = \frac{U}{d}\frac{Z}{X^2+Y^2+Z^2}Z = mZ. \quad (2)$$

Here $U$ is the electric voltage applied and $d$ the thickness of an electrooptic cell. In the spherical coordinate system ($X = r\sin\Theta\cos\varphi$, $Y = r\sin\Theta\sin\varphi$ and $Z = r\cos\Theta$) Eqs. (2) reduce to

$$E_1 = \frac{U}{d}\frac{\tan\Theta}{1+\tan^2\Theta}\cos\varphi, \quad (3)$$

$$E_2 = \frac{U}{d}\frac{\tan\Theta}{1+\tan^2\Theta}\sin\varphi, \quad (4)$$

$$E_3 = \frac{U}{d(1+\tan^2\Theta)}. \quad (5)$$

Such a field distribution can be produced using circular electrodes with different diameters (see Fig. 1).

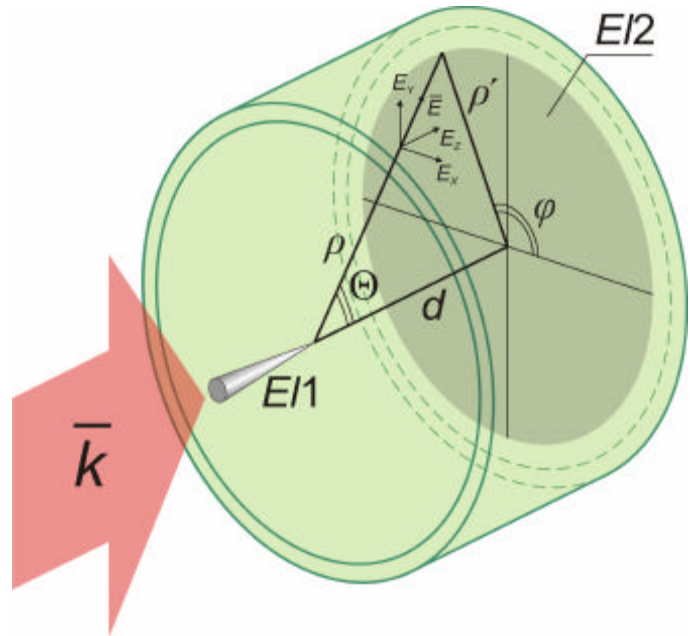

FIG. 1. Schematic representation of electrooptic cell with circular electrodes $El$1 and $El$2, and a conical spatial distribution of electric field produced by these electrodes.

In what follows we will analyze the Kerr effect induced by such a field in centrosymmetric media of different symmetries for the case of wide circularly polarized Gaussian beam propagating along the axis of electric field cone. Moreover, we will also consider the symmetry groups for which the quadratic electrooptic Kerr effect is not masked by the linear Pockels effect, which can be significantly greater in magnitude. In order to generate the vortex, the light beam should propagate along the initially isotropic direction, which remains isotropic in the center of cross section of the beam after the electric field is applied. The latter can happen only in the following cases: (3.A) isotropic media, (3.B) cubic crystals belonging to the symmetry groups m3m and 432, provided that the electric field is applied along one of the crystallographic or three-fold symmetry axes, (3.C) cubic crystals belonging to the group m3, whenever the field is applied in the same way as in the case (3.B), and (4.A)–(4.F) optically uniaxial crystals and textures under the electric field applied along the optic axis. Below we will thoroughly analyze all of these cases separately.

### 3. ISOTROPIC MEDIA AND CUBIC CRYSTALS

*A. Isotropic media with the symmetries $\infty/\infty/mmm$ and $\infty/\infty 2$*

It is obvious that the coordinate system *XYZ* for the isotropic media of the symmetry groups $\infty/\infty/mmm$ and $\infty/\infty 2$ can be chosen in an arbitrary manner. Before proceeding to the following analysis, we note that the third-rank Pockels tensor is equal to zero for the isotropic acentric symmetry group $\infty/\infty 2$. From the other side, the fourth-rank Kerr tensor is the same for the groups of symmetry $\infty/\infty/mmm$ and $\infty/\infty 2$. This tensor includes only two independent components, $R_{11}$ and $R_{12}$, and the relation $R_{44} = R_{11} - R_{12}$ holds true. When the electric field defined by Eqs. (2) is applied, the cross section of the optical indicatrix by the plane $Z = 0$ for the symmetry groups mentioned above is as follows:

$$(B_1 + R_{11}E_1^2 + R_{12}E_2^2 + R_{12}E_3^2)X^2 + (B_1 + R_{12}E_1^2 + R_{11}E_2^2 + R_{12}E_3^2)Y^2 + 2(R_{11} - R_{12})E_1 E_2 XY = 1. \quad (6)$$

This optical indicatrix perturbed by the electric field corresponds in fact to homogeneous media. However, Eqs. (2)–(5) imply that a material sample under such a conically shaped electric field should become inhomogeneous. For simulating a so-called effective angle of optical indicatrix orientation $z_{kl}^{ef}$ and an effective phase difference $\Delta\Gamma_{kl}^{ef}$ for each elementary ray, one can use a well known Jones matrix approach, with dividing the sample by, e.g., $N_{max} = 100$ homogeneous layers perpendicular to the *Z* axis. In its turn, each of the layers is divided by $k \times l$ homogeneous elementary cells in the *XY* plane. In practice, we have used a division given by $k = 100$ and $l = 100$. The number *n* of the layers which should be taken into consideration depends on the $r'$ coordinate (the module in the polar coordinate system – see Fig. 1): namely, it decreases with increasing $r'$. Thus, we have $n = 1 \div N_{max}$. The birefringence induced along the *Z* axis in each of the homogeneous cells is determined as

$$\Delta n_{XY} = -\frac{1}{2}n^3(R_{11} - R_{12})(E_1^2 + E_2^2) = -\frac{1}{2}n^3 m^2 (R_{11} - R_{12})(X^2 + Y^2), \quad (7)$$

while the angle of optical indicatrix rotation around the center of the *XY* cross section (i.e., the center of beam cross section) is given by

$$\tan 2\mathbf{z}_Z = \frac{2E_1 E_2}{E_1^2 - E_2^2} = \frac{2XY}{X^2 - Y^2} = \tan 2\mathbf{j}, \text{ or } \mathbf{z}_Z = \mathbf{j}. \tag{8}$$

Eq. (7) is nothing but equation of cone with a zero (singular) birefringence at the cone apex, i.e. in the center of the XY cross section. Then we readily obtain the resulting Jones matrix:

$$J_{kl} = \begin{vmatrix} \left(e^{i\Delta\Gamma_{kl}^{ef}/2}\cos^2 \mathbf{z}_{kl}^{ef} + e^{-i\Delta\Gamma_{kl}^{ef}/2}\sin^2 \mathbf{z}_{kl}^{ef}\right) & i\sin(\Delta\Gamma_{kl}^{ef}/2)\sin 2\mathbf{z}_{kl}^{ef} \\ i\sin(\Delta\Gamma_{kl}^{ef}/2)\sin 2\mathbf{z}_{kl}^{ef} & \left(e^{i\Delta\Gamma_{kl}^{ef}/2}\sin^2 \mathbf{z}_{kl}^{ef} + e^{-i\Delta\Gamma_{kl}^{ef}/2}\cos^2 \mathbf{z}_{kl}^{ef}\right) \end{vmatrix}$$
$$= \prod_{n=1}^{N_{\max}} \begin{vmatrix} \left(e^{i\Delta\Gamma_{kl}^{n}/2}\cos^2 \mathbf{z}_{kl}^{n} + e^{-i\Delta\Gamma_{kl}^{n}/2}\sin^2 \mathbf{z}_{kl}^{n}\right) & i\sin(\Delta\Gamma_{kl}^{n}/2)\sin 2\mathbf{z}_{kl}^{n} \\ i\sin(\Delta\Gamma_{kl}^{n}/2)\sin 2\mathbf{z}_{kl}^{n} & \left(e^{i\Delta\Gamma_{kl}^{n}/2}\sin^2 \mathbf{z}_{kl}^{n} + e^{-i\Delta\Gamma_{kl}^{n}/2}\cos^2 \mathbf{z}_{kl}^{n}\right) \end{vmatrix}, \tag{9}$$

where $\mathbf{DG}_{kl}^n = \frac{2\mathbf{p} d_{kl}^n}{\mathbf{l}}\left\{-\frac{1}{2}n^3\left(\frac{U}{d}\frac{Z}{X^2+Y^2+Z^2}\right)^2 (R_{11}-R_{12})(X^2+Y^2)\right\}$ and $\mathbf{z}_{kl}^n = \frac{1}{2}\arctan\frac{X}{Y}$

denote respectively the phase difference and the angle of optical indicatrix rotation within the elementary cells, and $d_{kl}^n$ is the thickness of the cell along the direction of light propagation.

We have simulated the effective phase difference $\mathbf{DG}_{kl}^{ef}$ and the spatial distribution of the optical indicatrix parameter $\mathbf{z}_{kl}^{ef}$ for a Kerr cell filled by a nitrobenzene liquid (the chemical formula $C_6H_5NO_2$, the effective Kerr coefficient $R_{11}-R_{12}=7.6\times 10^{-19}\,\text{m}^2/\text{V}^2$, and the refractive index $n=1.5562$ – see [25, 26]). It is seen from Fig. 2 (a, b) that the angle of optical indicatrix rotation $\mathbf{z}_{kl}^{ef}$ around the center of a wide beam depends linearly on the tracing angle $\mathbf{j}$. This should inevitably lead to appearance of a double-charged vortex whenever the sample is placed in between the crossed circular polarizers (the corresponding optical scheme is presented in Fig. 3). The appropriate relation is as follows:

$$\mathcal{E}^{out}(X,Y) = \mathcal{E}_0 \cos\frac{\mathbf{DG}_{kl}^{ef}}{2}\begin{bmatrix}1\\ \pm i\end{bmatrix} + i\mathcal{E}_0 \sin\frac{\mathbf{DG}_{kl}^{ef}}{2} e^{\pm i2p\mathbf{j}\pm i2\mathbf{z}_0}\begin{bmatrix}1\\ \mp i\end{bmatrix}, \tag{10}$$

where $?\mathbf{G}_{kl}^{ef}$ is the phase difference, $m=\pm 2p$ the topological charge of the vortex, $p$ the strength of topological defect, $\mathbf{z}_0$ the angle of optical indicatrix orientation at $\mathbf{j}=0$, and $\mathcal{E}_0$ and $\mathcal{E}^{out}(X,Y)$ denote the amplitudes of the incident and outgoing light wave, respectively (see, e.g., [27]). The second term in the r. h. s. of Eq. (10) describes the helical mode (the charge of the optical vortex is equal to two at $p=1$), while the first term describes the incident wave with no dislocation of the wave front present (it vanishes at $\Delta\Gamma_{kl}^{ef}=\mathbf{p}$). The induced phase difference is equal to zero in the center of the XY cross section, it increases with increasing distance from the center, and acquires its maximum when $r'_{\max}=1.96\,\text{mm}$ (see Fig. 2(c)). The birefringence fol-lows to zero with further increase of the module $r'$, since the electric field vanishes outside the field cone.

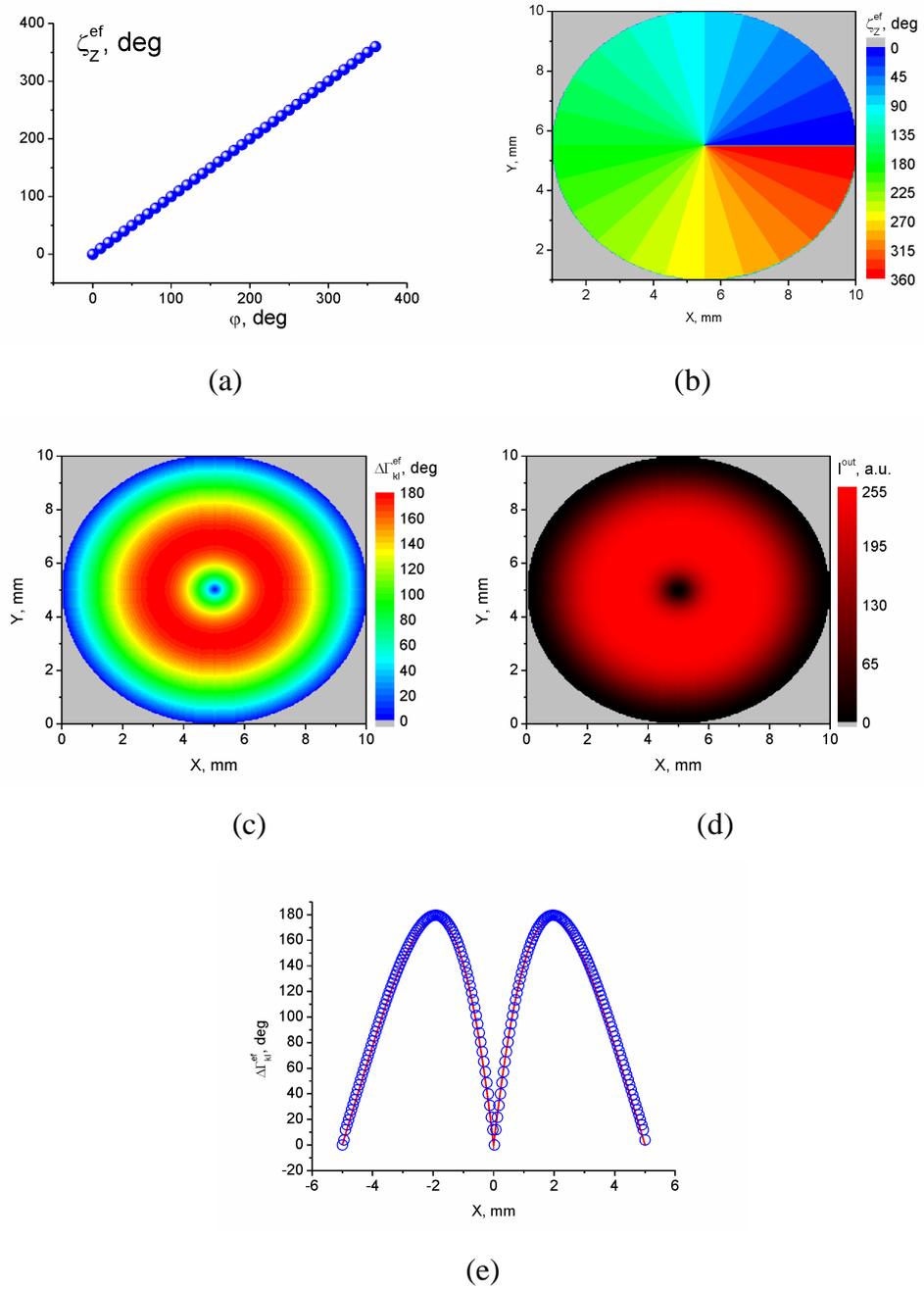

FIG. 2. (a) Calculated dependence of the effective angle of optical indicatrix rotation on the tracing angle at the constant module $r'$; (b) spatial *XY* distribution of the effective angle of optical indicatrix rotation; (c) induced effective phase difference; (d) appearance of the doughnut mode; and (e) dependence of the effective phase difference on the coordinate *X*, as calculated following from the Jones matrix approach (open circles) and formula (11) (solid curve). A nitrobenzene Kerr cell is assumed under the conical electric field $U/d = 1.98 \times 10^7$ V/m ($d = 5$ mm, the radius of the *El*2 electrode $R = 5$ mm, and the light wavelength $l = 632.8\,\text{nm}$).

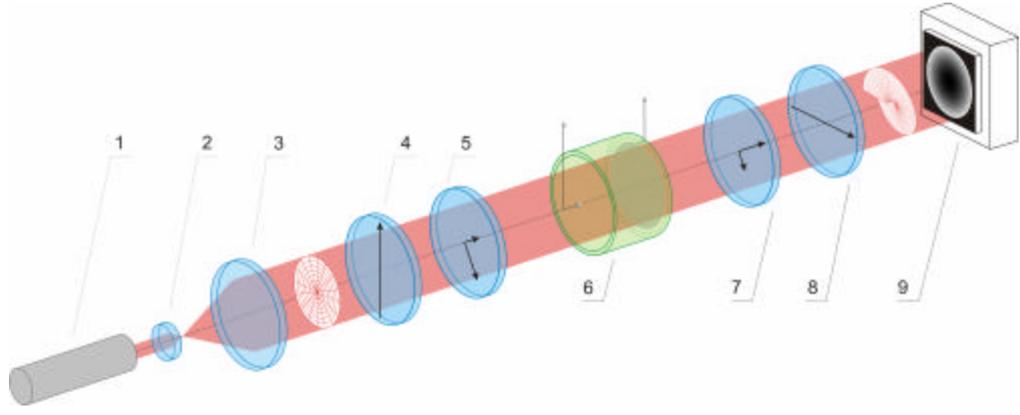

FIG. 3. Optical scheme of vortex generation: 1 – light source (e.g., laser), 2, 3 – objective lenses, 4, 8 – linear polarizers, 5, 7 – quarter-wave plates, 6 – sample (cuvette with a Kerr liquid), and 9 – CCD camera.

The effective phase difference can also be determined by integrating the local phase difference over the optical path ( $D G_{kl}^{ef} = \frac{2\pi}{\lambda} \int_{\frac{r'd}{R}}^{d} \Delta n_{XY} dZ$ ). Considering Eq. (7), one obtains

$$DG_{kl}^{ef} = \frac{\pi n^3 U^2 (R_{11} - R_{12}) r'}{2\lambda d^2} \left[ \frac{d(R(r'^2 + d^2) - r'(R^2 + d^2))}{(R^2 + d^2)(r'^2 + d^2)} + \arctan\left(\frac{d(R - r')}{r'R - d^2}\right) \right]. \quad (11)$$

We have verified that the dependences of the effective phase difference on the coordinate obtained basing on the Jones matrix approach and Eq. (11) yield in the same result (see Fig. 2(e)).

Spatial distribution of the light intensity in the doughnut mode can be simulated with the same Jones matrix technique. Let $\mathcal{E}_1$, $\mathcal{E}_2$ and $\mathcal{E}_1^{kl}$, $\mathcal{E}_2^{kl}$ be the components of the input and output Jones vectors, respectively, and $J^{QWP-}$, $J^{QWP+}$ and $J^A$ be the Jones matrices of the two quarter-wave plates rotated by 90 deg and the analyzer, respectively. Then we will get

$$\begin{pmatrix} \mathcal{E}_1^{kl} \\ \mathcal{E}_2^{kl} \end{pmatrix} = J^A J^{QWP-} J^{kl} J^{QWP+} \begin{pmatrix} \mathcal{E}_1 \\ \mathcal{E}_2 \end{pmatrix}, \quad (12)$$

where

$$\mathcal{E}_1 = 1, \quad \mathcal{E}_2 = 0, \quad J^A = \begin{pmatrix} 0 & 0 \\ 0 & 1 \end{pmatrix},$$

$$J^{QWP-} = \begin{pmatrix} \frac{1}{\sqrt{2}} e^{i\frac{\pi}{4}} & \frac{1}{\sqrt{2}} e^{-i\frac{\pi}{4}} \\ \frac{1}{\sqrt{2}} e^{-i\frac{\pi}{4}} & \frac{1}{\sqrt{2}} e^{i\frac{\pi}{4}} \end{pmatrix}, \quad J^{QWP+} = \begin{pmatrix} \frac{1}{\sqrt{2}} e^{-i\frac{\pi}{4}} & \frac{1}{\sqrt{2}} e^{i\frac{\pi}{4}} \\ \frac{1}{\sqrt{2}} e^{i\frac{\pi}{4}} & \frac{1}{\sqrt{2}} e^{-i\frac{\pi}{4}} \end{pmatrix}. \quad (13)$$

The resulting intensity for each of the elementary rays is determined by

$$\left(I^{kl}\right)^{out}_l = \begin{pmatrix} \mathcal{E}^{kl}_1 \\ \mathcal{E}^{kl}_2 \end{pmatrix} \begin{pmatrix} \mathcal{E}^{kl*}_1 \\ \mathcal{E}^{kl*}_2 \end{pmatrix}. \tag{14}$$

The corresponding spatial distribution of the intensity in the vortex mode induced by the electric field $U/d = 1.98 \times 10^7$ V/m in the nitrobenzene is presented in Fig. 2 (d).

### B. Cubic crystals of symmetry groups m3m and 432

In the crystals belonging to point symmetry groups m3m and 432, the third-rank tensor describing the linear electrooptic effect is equal to zero. For these materials the Kerr birefringence and the angle of optical indicatrix rotation induced by the conical electric field applied along the Z axis within the homogeneous cell are given by

$$\Delta n_{XY} = -\frac{1}{2} n^3 \sqrt{(R_{11} - R_{12})^2 (E_1^2 - E_2^2)^2 + 4 R_{44}^2 E_1^2 E_2^2}$$
$$= -\frac{1}{2} n^3 m^2 \sqrt{(R_{11} - R_{12})^2 (X^2 - Y^2)^2 + 4 R_{44}^2 X^2 Y^2}, \tag{15}$$

$$\tan 2\mathbf{z}_Z = \frac{2 R_{44} E_1 E_2}{(R_{11} - R_{12})(E_1^2 - E_2^2)} = \frac{2 R_{44} XY}{(R_{11} - R_{12})(X^2 - Y^2)} = \frac{R_{44}}{R_{11} - R_{12}} \tan 2\mathbf{j} . \tag{16}$$

It is clear that Eq. (15) represents an equation of hyperbolic cylinder. It can be transformed into equation of cone only under the condition $R_{44} = R_{11} - R_{12}$. Then the birefringence reads as $\Delta n_{XY} = -\frac{1}{2} n^3 m^2 (R_{11} - R_{12})(X^2 + Y^2)$, while the angle of optical indicatrix rotation becomes equal to the tracing angle (i.e., $\mathbf{z}_Z = \mathbf{j}$ ). As a consequence, one has to observe a canonical double-charged vortex whenever the sample is placed in between the circular polarizers. This case matches with the case (3.A) considered above for the isotropic media.

Now let us analyze what will happen when $R_{44} \neq R_{11} - R_{12}$. Using Eq. (9), one can simulate the effective angle of optical indicatrix rotation and the effective phase difference for the ratio $R_{44}/(R_{11} - R_{12})$ equal, e.g., to 0.1. One can see from Fig. 4 (a, b) that the both spatial distributions are divided into four parts. The dependence of the angle of optical indicatrix rotation has a step-like character (see Fig. 4 (a)). As follows from Eq. (10), the phase of the helical mode is determined by either $\Gamma = \pm 2p\mathbf{j} \pm 2\mathbf{z}_0 = \pm 2\mathbf{z} \pm 2\mathbf{z}_0$ or $\Delta\Gamma = 2\mathbf{z}$. Thus, the phase increases twice as faster than the angle of optical indicatrix rotation with changing tracing angle. The dependences of the phase of the doughnut mode on the tracing angle for different ratios $R_{44}/(R_{11} - R_{12})$ are presented in Fig. 4 (c). With increasing ratio $R_{44}/(R_{11} - R_{12})$, beginning from its unit value, a pure screw dislocation of the phase front gradually transforms into a pure edge dislocations of the phase front. Notice that in this case we deal with two edge dislocations crossed at the angle of 90 deg. For the intermediate ratios, a mixed screw-edge dislocation has to be observed.

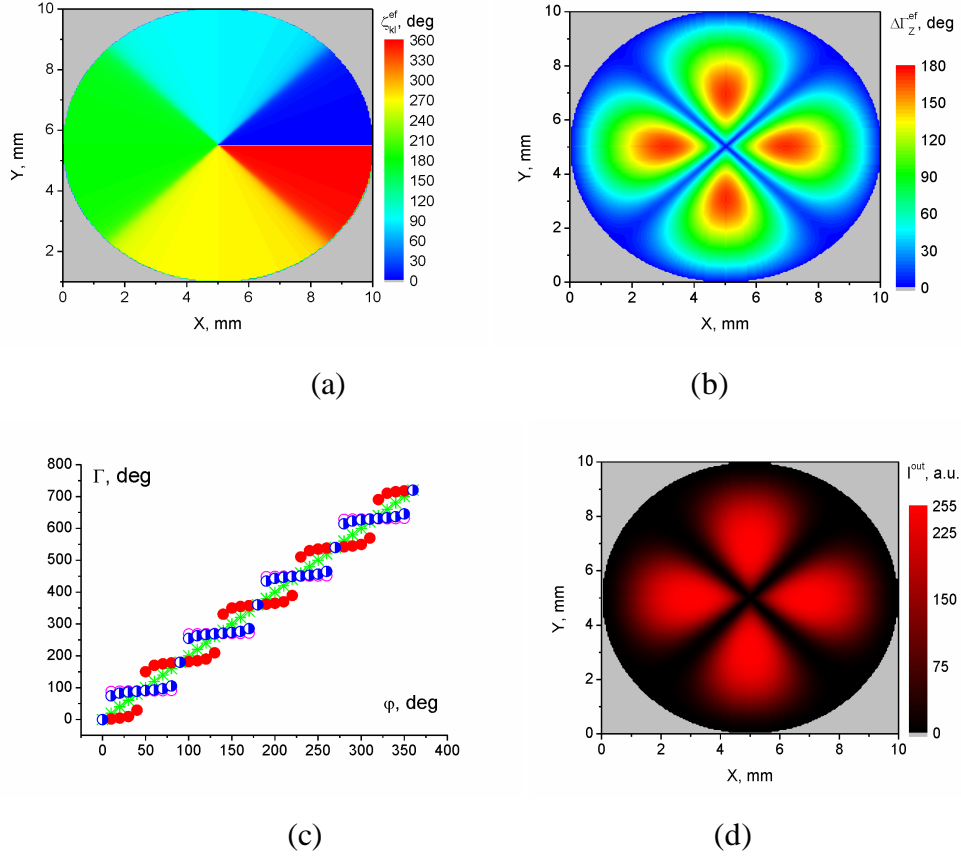

(a) (b)

(c) (d)

FIG. 4. (a) Calculated *XY* distribution of the effective angle of optical indicatrix rotation at $R_{44}/(R_{11}-R_{12})=0.1$; (b) induced effective phase difference at $R_{44}/(R_{11}-R_{12})=0.1$ ($R_{11}=1.09\times10^{-18}$ m$^2$/V$^2$, $R_{12}=0.29\times10^{-18}$ m$^2$/V$^2$ and $R_{44}=0.08\times10^{-18}$ m$^2$/V$^2$) under the conical electric field $U/d=16\times10^6$ m/V; (c) dependence of the phase of outgoing light beam on the tracing angle at different ratios $R_{44}/(R_{11}-R_{12})$ (1 – stars, 0.1 – full circles, 10 – semi-open circles, and 100 – open circles); and (d) intensity distribution for the outgoing beam at $R_{44}/(R_{11}-R_{12})=0.1$ ($d=5$ mm, $R=5$ mm and $l=632.8$ nm – see the notation in Fig. 2).

Let us illustrate the effect mentioned above using the parameters of real crystals in our simulations, e.g., SrTiO$_3$ belonging to the point symmetry group m3m. It is characterized by the Kerr coefficients $R_{11}=1.09\times10^{-18}$ m$^2$/V$^2$, $R_{12}=0.29\times10^{-18}$ m$^2$/V$^2$ and $R_{44}=0.58\times10^{-18}$ m$^2$/V$^2$, and the refractive index $n=2.38$ at $l=632.8$ nm [28]. Therefore the ratio of the Kerr coefficients for the SrTiO$_3$ crystals is equal to $R_{44}/(R_{11}-R_{12})=0.73$, which is quite close to unity. Fig. 5 shows spatial distributions of the effective phase difference and the angle of optical indicatrix rotation in the *XY* cross section, and appearance of the doughnut mode in case when the conical electric field $U/d = 1.01 \times 10^7$ V/m is applied and the light beam propagates along the *Z* direction. As seen from Fig. 5, the mixed screw-edge dislocation can really appear in SrTiO$_3$, while the doughnut mode should be slightly fragmentized. In fact, the practical conditions provided by the SrTiO$_3$ crystals are very close to the case of generation of the pure screw dislocation and, as seen from Fig. 5 (c), the doughnut mode is almost circular.

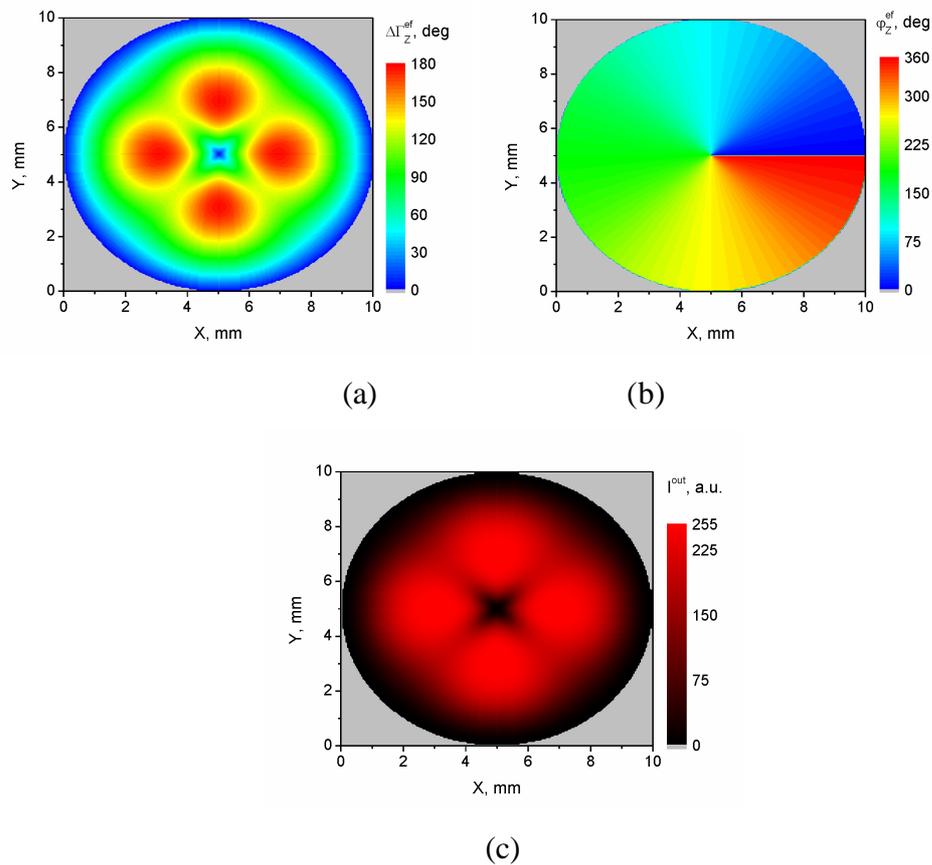

FIG. 5. Calculated distributions of the effective phase difference (a) and the effective angle of optical indicatrix rotation (b), and light intensity behind the circular analyzer (c). A conical electric field $U/d = 1.01 \times 10^7$ V/m is applied along the *Z* direction in SrTiO$_3$ crystals ($d$ = 5 mm, $R$ = 5 mm, and $\lambda$ = 632.8 nm – see the notation in Fig. 2).

Now let us consider propagation of a wide Gaussian beam, assuming that a conical electric field is applied along the direction [111] in cubic crystals of the symmetry groups m3m and 432. Then it would be convenient to rewrite the fourth-rank Kerr tensor in the coordinate system of which $Z'$ axis is parallel to the direction [111], $X'$ to [11$\bar{2}$], and $Y'$ to [$\bar{1}$10]. The tensor of the Kerr coefficients becomes

|        | $E'_1 E'_1$              | $E'_2 E'_2$              | $E'_3 E'_3$ | $E'_2 E'_3$ | $E'_1 E'_3$ | $E'_1 E'_2$ |
|--------|--------------------------|--------------------------|-------------|-------------|-------------|-------------|
| $\Delta B'_1$ | $R'_{11}$ | $R'_{12}$ | $R'_{13}$ | $R'_{14}$ | $-R'_{14}$ | $0$ |
| $\Delta B'_2$ | $R'_{12}$ | $R'_{11}$ | $R'_{13}$ | $-R'_{14}$ | $R'_{14}$ | $0$ |
| $\Delta B'_3$ | $R'_{13}$ | $R'_{13}$ | $R'_{33}$ | $0$ | $0$ | $0$ |
| $\Delta B'_4$ | $\frac{1}{2}R'_{14}$ | $-\frac{1}{2}R'_{14}$ | $0$ | $R'_{44}$ | $0$ | $R'_{14}$ |
| $\Delta B'_5$ | $-\frac{1}{2}R'_{14}$ | $\frac{1}{2}R'_{14}$ | $0$ | $0$ | $R'_{44}$ | $R'_{14}$ |
| $\Delta B'_6$ | $0$ | $0$ | $0$ | $R'_{14}$ | $R'_{14}$ | $R'_{66}$ |

(17)

with the components $R'_{11} = \frac{1}{2}(R_{11} + R_{44} + R_{12})$, $R'_{12} = \frac{1}{6}(R_{11} - R_{44}) + \frac{5}{6}R_{12}$, $R'_{13} = \frac{1}{3}(R_{11} - R_{44} + 2R_{12})$, $R'_{14} = \frac{1}{3}(R_{11} - R_{44} - R_{12})$, $R'_{33} = \frac{1}{3}(R_{11} + 2R_{44} + 2R_{12})$, $R'_{44} = \frac{1}{3}(2R_{11} + R_{44} - 2R_{12})$, and $R'_{66} = \frac{1}{3}(R_{11} + 2R_{44} - R_{12})$. The birefringence induced along the $Z'$ direction and the rotation of optical indicatrix around the same direction within one elementary cell may be written as

$$\begin{aligned}\Delta n_{X'Y'} &= -\frac{n^3}{2}\left(\begin{array}{l}\left[R'_{66}(E'^2_1 - E'^2_2) - 2R'_{14}E'_3(E'_1 - E'_2)\right]^2 \\ +4\left[R'_{14}E'_3(E'_1 + E'_2) + R'_{66}E'_1E'_2\right]^2\end{array}\right)^{1/2} \\ &= -\frac{n^3}{6}\left(\begin{array}{l}\left[(R_{11} + 2R_{44} - R_{12})(E'^2_1 - E'^2_2) - 2(R_{11} - R_{44} - R_{12})E'_3(E'_1 - E'_2)\right]^2 \\ +4\left[(R_{11} - R_{44} - R_{12})E'_3(E'_1 + E'_2) + (R_{11} + 2R_{44} - R_{12})E'_1E'_2\right]^2\end{array}\right)^{1/2}\end{aligned}$$ (18)

and

$$\begin{aligned}\tan 2z_{Z'} &= \frac{2(R'_{14}E'_3(E'_1 + E'_2) + R'_{66}E'_1E'_2)}{R'_{66}(E'^2_1 - E'^2_2) - 2R'_{14}E'_3(E'_1 - E'_2)} \\ &= \frac{2((R_{11} - R_{44} - R_{12})E'_3(E'_1 + E'_2) + (R_{11} + 2R_{44} - R_{12})E'_1E'_2)}{(R_{11} + 2R_{44} - R_{12})(E'^2_1 - E'^2_2) - 2(R_{11} - R_{44} - R_{12})E'_3(E'_1 - E'_2)},\end{aligned}$$ (19)

respectively. These relations include the electric field component $E'_3$. Let us remind that this component is not equal to zero at the axis of the electric field cone (see Eqs. (2)–(5)), where it acquires its maximum value. Availability of the $E'_3$ component can lead to nonzero birefringence in the center of the XY cross section and appearance of a nonzero light intensity in the vortex nucleus. However, the component $E'_3$ appearing in Eqs. (18) and (19) is always multiplied by the components $E'_1$ and $E'_2$, which become zero at the axis of the cone of electric field.

Notice that, under the condition $R_{44} = R_{11} - R_{12}$, Eqs. (18) and (19) can be represented as

$$\Delta n_{X'Y'} = -\frac{n^3}{2}R_{44}(E'^2_1 + E'^2_2) = -\frac{n^3}{2}R_{44}m^2(X'^2 + Y'^2),$$ (20)

$$\tan 2\mathbf{z}_{Z'} = \frac{2E'_1 E'_2}{E'^2_1 - E'^2_2} = \frac{2X'Y'}{X'^2 - Y'^2} = \tan 2\mathbf{j}, \text{ or } \mathbf{z}_{Z'} = \mathbf{j}. \tag{21}$$

Eqs. (20) and (21) are similar to Eqs. (7) and (8) describing the polarization singularity that leads to the appearance of a canonical double-charged vortex.

In fact, one can analyze two extreme conditions when considering Eq. (19). The first one corresponds to $R'_{14} = 0$ (leading to vanishing $E'_3$ component), which leads to Eq. (21) and appearance of the double-charged vortex. The second corresponds to $R'_{66} = 0$. In the latter case it follows from Eq. (19) that $\tan 2\mathbf{z}_{Z'} = \tan(\mathbf{p}/4 + \mathbf{j})$ (or $\mathbf{z}_{Z'} = \mathbf{p}/8 + \mathbf{j}/2$), i.e. the single-charged vortex has to appear. Then the following question appears: will the charge of the vortex embedded into the emergent beam be non-integer when $R'_{66}, R'_{14} \neq 0$? Such a situation could correspond to appearance of vortices with fractional topological charges (see [29-32]).

The spatial distributions of the phase difference and the angle of optical indicatrix rotation, as well as the dependence of intensity of the emergent beam on the ratio $\frac{R'_{14}}{R'_{66}} = \frac{(R_{11} - R_{12} - R_{44})}{(R_{11} - R_{12} + 2R_{44})}$ are presented in Fig. 6. The denominator in the ratio mentioned above is always greater than the numerator, since $2R_{44} > -R_{44}$. Therefore we have $0 \leq R'_{14}/R'_{66} < 1$ and the pure single-charged vortex cannot in principle be obtained under these conditions. As seen from Fig. 6 (a), the conical distribution of the phase difference is produced at $R'_{14}/R'_{66} = 0$. With this ratio, the strength of the topological defect of the optical indicatrix orientation is equal to +1 (see Fig. 6 (e, m)). This corresponds to generation of the double-charged vortex in the center of the beam cross section, or the doughnut mode with the OAM per photon equal to $2\hbar$ (see Fig. 6 (i)) [33, 34].

With the ratio $R'_{14}/R'_{66} = 0.01$, the spatial distribution of the phase difference and the intensity of the emergent light become inhomogeneous, in relation to dependence on the tracing angle (see Fig. 6 (b, j)). In fact, they become triangular, while the vortex beam gets distorted. Nonetheless, the strength of the topological defect of the optical indicatrix orientation remains to be equal +1, thus corresponding to generation of the OAM per photon equal to $|2\hbar|$ (see Fig. 6 (f, j, n)). Notice that similar triangular shapes of the phase difference and the intensity distribution are peculiar for the ratio $R'_{14}/R'_{66} = 0.99$ (see Fig. 6 (d, l)). However, in this case the strength of the topological defect of the optical indicatrix orientation is equal to $-1/2$ (see Fig. 6 (p)). This corresponds to generation of the single-charged vortex and the helical mode with the OAM per photon equal to $|\hbar|$ (see Fig. 6 (h)).

In fact, five polarization singularities appear at $R'_{14}/R'_{66} = 0.25$, as seen from Fig. 6 (c, g, o) with the strength of topological defect of the optical indicatrix orientation equal to $\pm 1/2$. Two of them has the strength $+1/2$ indicating that the supposed fractional vortex charge is close to two, whereas the three remaining ones represents the chain with alternating sign of the strengths of defects $\pm 1/2$ (see Fig. 6 (o) and notice also that the OAM of each of the vortices is equal to $|\hbar|$). Hence, five single-charged vortices have to appear under such conditions. Nonetheless, no vortices with fractional charges are observed. This fact agrees with the results [32] showing that only the vortices with integer charges should appear when a spiral phase plate is used with a step which is not aliquot to the light wavelength.

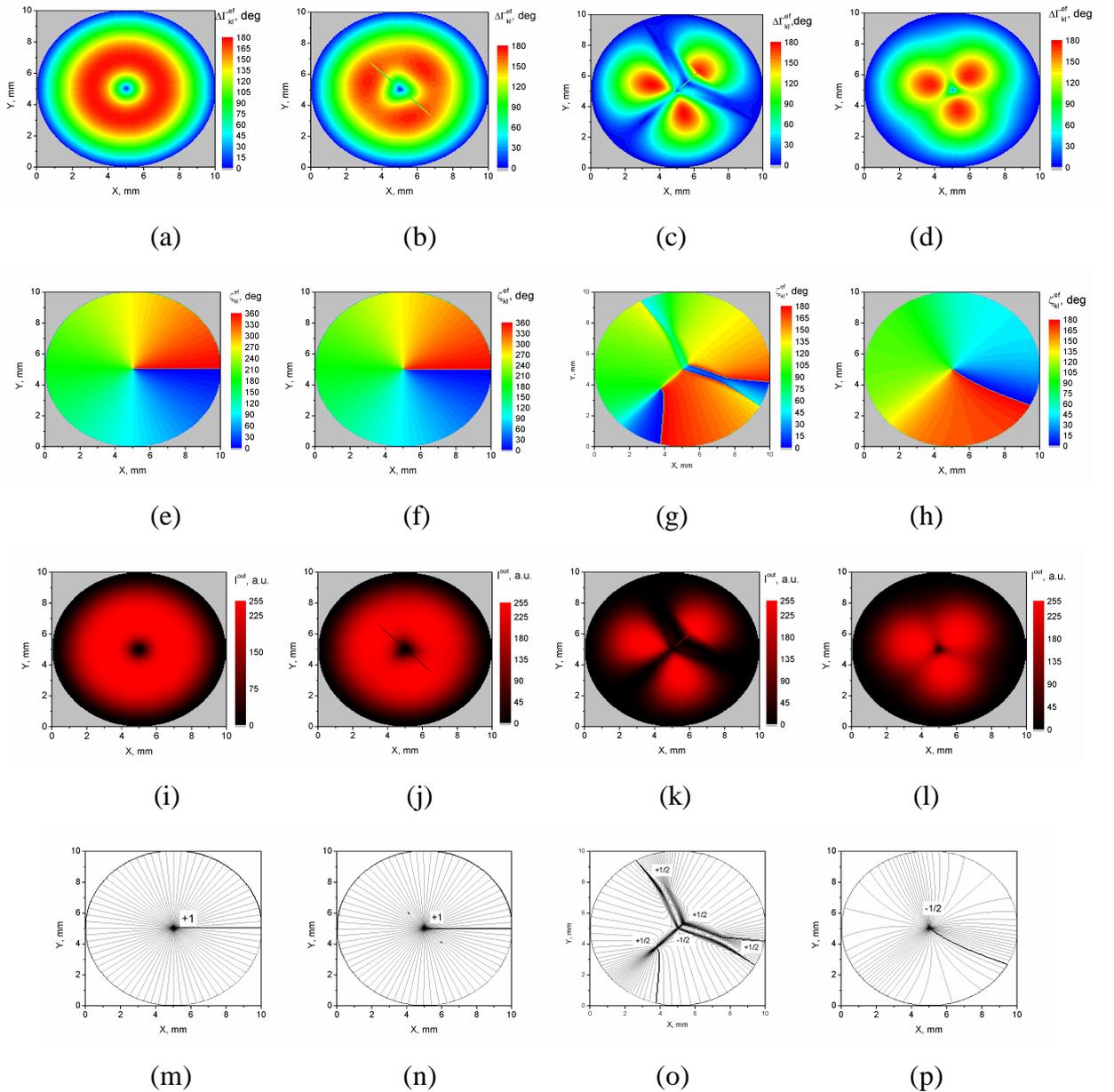

FIG. 6. Calculated spatial distributions of effective phase difference (a–d) and effective angle of optical indicatrix rotation (e–h); light intensity behind a circular analyzer (i–l) and schemes of topological defects (m–p) induced by a conical electric field applied along the direction [111] in crystals belonging to the point symmetry groups m3m and 432. The ratio is $R'_{14}/R'_{66} = 0$ (a, e, i, m), $U = 45$ kV, $R'_{14}/R'_{66} = 0.01$ (b, f, j, n), $U = 43.5$ kV and $R'_{14}/R'_{66} = 0.25$ (c, g, k, o), and $U = 30$ kV and $R'_{14}/R'_{66} = 0.99$ (d, h, l, p) at $U = 18$ kV.

### C. Cubic crystals of symmetry group m3

For the group of symmetry m3 we have $R_{12} \neq R_{21}$ and $R_{44} \neq R_{11} - R_{12}$. When a conical electric field is applied along the [001] axis, the optical indicatrix of the homogeneous cell is given by the formula

$$(B_1 + R_{11}E_1^2 + R_{12}E_2^2 + R_{21}E_3^2)X^2 + (B_1 + R_{21}E_1^2 + R_{11}E_2^2 + R_{12}E_3^2)Y^2 + 2R_{44}E_1E_2 XY = 1. \tag{22}$$

Then the induced birefringence and the angle of optical indicatrix rotation within the homogeneous cell are described by the relations

$$\Delta n_{XY} = -\frac{1}{2}n^3 \sqrt{\left[(R_{11} - R_{21})E_1^2 - (R_{11} - R_{12})E_2^2 + (R_{21} - R_{12})E_3^2\right]^2 + 4R_{44}^2 E_1^2 E_2^2}$$
$$= -\frac{1}{2}n^3 m^2 \sqrt{\left[(R_{11} - R_{21})X^2 - (R_{11} - R_{12})Y^2 + (R_{21} - R_{12})Z^2\right]^2 + 4R_{44}^2 X^2 Y^2}, \tag{23}$$

$$\tan 2z_Z = \frac{2R_{44}E_1 E_2}{(R_{11} - R_{21})E_1^2 - (R_{11} - R_{12})E_2^2 + (R_{21} - R_{12})E_3^2}$$
$$= \frac{2R_{44} XY}{(R_{11} - R_{21})X^2 - (R_{11} - R_{12})Y^2 + (R_{21} - R_{12})Z^2}. \tag{24}$$

Only in case when $R_{12} = R_{21}$ and $R_{44} = R_{11} - R_{12}$, these relations may be simplified to

$$\Delta n_{XY} = -\frac{1}{2}n^3 R_{44}(E_1^2 + E_2^2) = -\frac{1}{2}n^3 m^2 R_{44}(X^2 + Y^2) \tag{25}$$

and

$$\tan 2z_Z = \frac{2E_1 E_2}{E_1^2 - E_2^2} = \frac{2XY}{X^2 - Y^2} = \tan 2j \text{ or } z_Z = j. \tag{26}$$

These equations describe a polarization singularity which can be transformed to the canonical double-charged vortex (see, e.g., Fig. 2). In case when $R_{12} = R_{21}$ and $R_{44} \neq R_{11} - R_{12}$, Eqs. (23) and (24) may be rewritten as

$$\Delta n_{XY} = -\frac{1}{2}n^3 \sqrt{(R_{11} - R_{12})^2 (E_1^2 - E_2^2)^2 + 4R_{44}^2 E_1^2 E_2^2}$$
$$= -\frac{1}{2}n^3 m^2 \sqrt{(R_{11} - R_{12})(X^2 - Y^2)^2 + 4R_{44}^2 X^2 Y^2} \tag{27}$$

and

$$\tan 2\mathbf{z}_Z = \frac{R_{44}}{(R_{11}-R_{12})}\frac{2E_1E_2}{(E_1^2-E_2^2)} = \frac{R_{44}}{(R_{11}-R_{12})}\frac{2XY}{(X^2-Y^2)} = \frac{R_{44}}{R_{11}-R_{12}}\tan 2\mathbf{j}, \qquad (28)$$

respectively. These relations are the same as Eqs. (15) and (16), which describe a polarization singularity leading to a mixed screw-edge dislocation of the wave front (see Fig. 4).

In case when the conical electric field is applied along the direction [111], one can apply a procedure similar to that mentioned above, i.e. to rewrite the Kerr tensor in the coordinate system of which $Z'$ axis is parallel to the direction [111], $X'$ to [11$\bar{2}$], and $Y'$ to [$\bar{1}$10]. However, this time the tensor is quite complicated and includes 12 nonzero components, of which 8 are independent. Taking into account that $R'_{66} = R'_{11} - R'_{12}$, one can find the optical birefringence and the angle of optical indicatrix rotation:

$$\Delta n_{X'Y'} = -\frac{n^3}{2}\left[\begin{array}{l}\left(R'_{66}(E_1'^2-E_2'^2)+2R'_{14}E'_3E'_2+2R'_{15}E'_3E'_1+2R'_{16}E'_1E'_2\right)^2 \\ +4\left(\frac{1}{2}R'_{16}(E_2'^2-E_1'^2)-R'_{15}E'_3E'_2+R'_{14}E'_3E'_1+R'_{66}E'_1E'_2\right)^2\end{array}\right]^{1/2}, \qquad (29)$$

$$\tan 2\mathbf{z}_{Z'} = \frac{R'_{16}(E_2'^2-E_1'^2)-2R'_{15}E'_3E'_2+2R'_{14}E'_3E'_1+2R'_{66}E'_1E'_2}{R'_{66}(E_1'^2-E_2'^2)+2R'_{14}E'_3E'_2+2R'_{15}E'_3E'_1+2R'_{16}E'_1E'_2}. \qquad (30)$$

The Kerr coefficients $R'_{lm}$ appearing in these relations represent components of the tensor rewritten in the new coordinate system $X'Y'Z'$. As shown above, the features of the $E'_3$ component of the electric field lead to multiplication of the polarization singularities, with appearance of a number of single-charged vortices and mixed screw-edge dislocations of the wave front.

The conditions under which this field component vanishes can be formulated as $R'_{14} = 0$ and $R'_{15} = 0$, where

$$R'_{14} = \frac{1}{3}(R_{11}-R_{44}) - \frac{1+\sqrt{3}}{6}R_{12} - \frac{1-\sqrt{3}}{6}R_{21}, \quad R'_{15} = -\frac{1}{3}(R_{11}-R_{44}) + \frac{1-\sqrt{3}}{6}R_{12} + \frac{1+\sqrt{3}}{6}R_{21}. \quad (31)$$

It is seen from the above equations that $R'_{14} = 0$ and $R'_{15} = 0$ when $R_{44} = R_{11} - R_{12}$ and $R_{12} = R_{21}$. Then Eqs. (29) and (30) reduce to

$$\Delta n_{X'Y'} = -\frac{n^3}{2}\sqrt{R_{16}'^2+R_{66}'^2}(E_1'^2+E_2'^2) = -\frac{n^3}{2}m^2\sqrt{R_{16}'^2+R_{66}'^2}(X'^2+Y'^2) \qquad (32)$$

$$\tan 2\mathbf{z}_{Z'} = \frac{R'_{16}(E_2'^2-E_1'^2)+2R'_{66}E'_1E'_2}{R'_{66}(E_1'^2-E_2'^2)+2R'_{16}E'_1E'_2} = \frac{\tan 2\mathbf{j} - \frac{R'_{16}}{R'_{66}}}{1+\frac{R'_{16}}{R'_{66}}\tan 2\mathbf{j}} = \tan(2\mathbf{j}-2\mathbf{z}_0), \quad \tan 2\mathbf{z}_0 = \frac{R'_{16}}{R'_{66}}. \quad (33)$$

It follows from these formulas that the $X'Y'$-distribution of the induced birefringence within the cell reveals a conical shape. The change in the angle of optical indicatrix rotation around the geometrical center of the $X'Y'$ cross section is the same as the tracing angle. Hence, the double-charged vortex would appear behind the circular analyzer only in case when $R_{44} = R_{11} - R_{12}$ and $R_{12} = R_{21}$, and the conical electric field is applied along the direction [111] in the crystals belonging to the point symmetry m3.

Summarizing the results presented above, we make a general conclusion that the pure double-charged vortex beam can be created using the Kerr effect only under the condition defined by the relation $R_{44} = R_{11} - R_{12}$.

## 4. OPTICALLY UNIAXIAL CRYSTALS AND TEXTURES

### A. Hexagonal crystals and textures of symmetry groups 622, 6mm, 6/mmm, ∞2, ∞mm and ∞/mmm

Despite the groups 622, 6mm, ∞2 and ∞mm do not include an inversion center among their symmetry operations, their Pockels coefficients do not affect the *XY* distribution of the induced birefringence. Let us notice that the relationship $R_{66} = R_{11} - R_{12}$ holds true for the symmetry groups 622, 6mm, 6/mmm, ∞2, ∞mm and ∞/mmm. When the conical field is applied along the *Z* direction, the birefringence and the angle of optical indicatrix rotation within the homogeneous cell may be written respectively as

$$\Delta n_{XY} = -\frac{1}{2}n_o^3(R_{11} - R_{12})(E_1^2 + E_2^2) = -\frac{1}{2}n_o^3 m^2 (R_{11} - R_{12})(X^2 + Y^2), \quad (34)$$

and

$$\tan 2z_Z = \tan 2j \text{ or } z_Z = j. \quad (35)$$

These relations are the same as Eqs. (6) and (7) for the isotropic media. Thus, the dependences presented in Fig. 2 will also be valid for all of the symmetry groups just considered. Then the conically shaped electric field will produce the canonical double-charged optical vortex.

### B. Hexagonal crystals of symmetry groups 6 and 6/m and textures of symmetries ∞/m and ∞

For the symmetry groups 6, 6/m, ∞/m and ∞, the Kerr coefficients are linked as $R_{66} = R_{11} - R_{12}$. Let us notice that the Pockels birefringence in the geometry under consideration is equal to zero for the acentric groups 6 and ∞. Then the induced birefringence and the angle of optical indicatrix rotation within the homogeneous cell may be presented by the formulas

$$\Delta n_{XY} = -\frac{1}{2}n_o^3 \sqrt{R_{66}^2 + 4R_{62}^2}(E_1^2 + E_2^2) = -\frac{1}{2}n_o^3 m^2 \sqrt{R_{66}^2 + 4R_{62}^2}(X^2 + Y^2) \quad (36)$$

and

$$\tan 2z_Z = \tan 2(j - z_0), \quad z_Z = j - z_0, \quad z_0 = \frac{1}{2}\arctan\frac{2R_{62}}{R_{66}}, \tag{37}$$

respectively. Taking the equality $R_{62} = -R_{61}$ into account, we rewrite these relations as

$$\Delta n_{XY} = -\frac{1}{2}n_o^3\sqrt{R_{66}^2 + 4R_{61}^2}(E_1^2 + E_2^2) = -\frac{1}{2}n_o^3 m^2 \sqrt{R_{66}^2 + 4R_{61}^2}(X^2 + Y^2), \tag{38}$$

$$\tan 2z_Z = \tan 2(j + z_0), \quad z_Z = j + z_0, \quad z_0 = \frac{1}{2}\arctan\frac{2R_{61}}{R_{66}}. \tag{39}$$

As a result, the *XY* distribution of the induced birefringence within the cell has a conical shape. The change in the angle of optical indicatrix rotation around the geometrical center of the *XY* cross section is the same as the tracing angle. Therefore the double-charged vortex should appear behind the circular analyzer. This vortex is the same as for the isotropic materials and for the crystals and textures of the group (4.A) (see Fig. 2). The only difference taking place in the present case is that the initial angle of optical indicatrix rotation is defined by the ratio $2R_{61}/R_{66}$ and so remains nonzero at $j = 0\deg$.

### C. Tetragonal crystals symmetry groups 422, 4mm, $\overline{4}2m$ and 4/mmm

The induced Pockels birefringence in the chosen interaction geometry is equal to zero for the acentric symmetry groups 422, 4mm and $\overline{4}2m$. As far as the groups 422, 4mm, $\overline{4}2m$ and 4/mmm are concerned, the birefringence and the angle of optical indicatrix rotation within the homogeneous cell result in

$$\begin{aligned}\Delta n_{XY} &= -\frac{1}{2}n_o^3\sqrt{(R_{11} - R_{12})^2(E_1^2 - E_2^2)^2 + 4R_{66}^2 E_1^2 E_2^2} \\ &= -\frac{1}{2}n_o^3 m^2 \sqrt{(R_{11} - R_{12})^2(X^2 - Y^2)^2 + 4R_{66}^2 X^2 Y^2}\end{aligned}, \tag{40}$$

$$\tan 2z_Z = \frac{R_{66}}{R_{11} - R_{12}}\tan 2j. \tag{41}$$

It is obvious that in this case a mixed screw-edge dislocation of the wave front appears. Only when $R_{66} = R_{11} - R_{12}$, it can be transformed into a pure screw dislocation, with a possibility for generating a canonical double-charged vortex.

### D. Tetragonal crystals of symmetry groups 4 and 4/m

The relations for the birefringence and the angle of optical indicatrix rotation for the groups of symmetry 4 and 4/m look as follows:

$$\Delta n_{XY} = -\frac{1}{2} n_o^3 \sqrt{\left[(R_{11} - R_{12})(E_1^2 - E_2^2) + 2R_{16}E_1E_2\right]^2 + 4\left[R_{61}(E_1^2 - E_2^2) + R_{66}E_1E_2\right]^2}$$

$$= -\frac{1}{2} n_o^3 m^2 \sqrt{\left[(R_{11} - R_{12})(X^2 - Y^2) + 2R_{16}XY\right]^2 + 4\left[R_{61}(X^2 - Y^2) + R_{66}XY\right]^2}, \quad (42)$$

$$\tan 2z_Z = \frac{2R_{61} + R_{66} \tan 2j}{(R_{11} - R_{12}) + R_{16} \tan 2j}. \quad (43)$$

Therefore a mixed screw-edge dislocation of the wave front is induced in these crystals. Only when $R_{11} - R_{12} = R_{66}$ and $R_{61} = R_{16} = 0$, Eq. (42) describes a canonical cone and can be transformed to

$$\Delta n_{XY} = -\frac{1}{2} n_o^3 m^2 R_{66} (X^2 + Y^2). \quad (44)$$

However, such a ratio among the Kerr coefficients seems to be unlikely.

### E. Trigonal crystals of symmetry group $\bar{3}$m

The symmetry group $\bar{3}$m is known for the relationship $R_{66} = R_{11} - R_{12}$, though the matrix of the Kerr coefficients includes an additional component $R_{14}$. Then the birefringence and the angle of optical indicatrix rotation for the homogeneous cell become as follows:

$$\Delta n_{XY} = -\frac{1}{2} n_o^3 \sqrt{R_{66}^2 (E_1^2 + E_2^2)^2 + 4R_{66}R_{14}E_2E_3(3E_1^2 - E_2^2) + 4R_{14}^2 E_3^2(E_1^2 + E_2^2)}, \quad (45)$$

$$\tan 2z_Z = \frac{2(R_{14}E_3E_1 + R_{66}E_1E_2)}{R_{66}(E_1^2 - E_2^2) + 2R_{14}E_2E_3}. \quad (46)$$

A conical distribution of the birefringence can be induced whenever $R_{14} = 0$. Then we get

$$\Delta n_{XY} = -\frac{1}{2} n_o^3 R_{66} (E_1^2 + E_2^2) = -\frac{1}{2} n_o^3 R_{66} m^2 (X^2 + Y^2), \quad (47)$$

$$\tan 2z_Z = \frac{2E_1E_2}{E_1^2 - E_2^2} = \tan 2j, \quad z_Z = j. \quad (48)$$

Otherwise Eqs. (45) and (46) would correspond to a number of single-charged vortices whose properties depend on the ratio $R_{14}/R_{66}$ (see Fig. 6).

### F. Trigonal crystals of symmetry group $\bar{3}$

Among all the symmetry groups considered above, the group $\bar{3}$ reveals the lowest symmetry. In spite of this fact, we have the link $R_{66} = R_{11} - R_{12}$. The relations for the induced birefringence and the angle of optical indicatrix rotation for the homogeneous cell become rather complicated:

$$\Delta n_{XY} = -\frac{1}{2}n_o^3 \sqrt{\begin{array}{l}\left[R_{66}(E_1^2 - E_2^2) + 2R_{14}E_2E_3 - 2R_{25}E_3E_1 + 4R_{62}E_1E_2\right]^2 \\ +4\left[R_{25}E_2E_3 + R_{14}E_3E_1 + R_{66}E_1E_2 - R_{62}(E_1^2 - E_2^2)\right]^2\end{array}}, \tag{49}$$

$$\tan 2\mathbf{z}_Z = \frac{2(R_{62}(E_2^2 - E_1^2) + R_{25}E_2E_3 + R_{14}E_3E_1 + R_{66}E_1E_2)}{R_{66}(E_1^2 - E_2^2) + 2R_{14}E_2E_3 - 2R_{25}E_3E_1 + 4R_{62}E_1E_2}. \tag{50}$$

Only in the case when $R_{25} = R_{14} = R_{62} = 0$, the relation for the birefringence may be reduced to the conical form

$$\Delta n_{XY} = -\frac{1}{2}n_o^3 R_{66}(E_1^2 + E_2^2) = -\frac{1}{2}n_o^3 R_{66}m^2(X^2 + Y^2), \tag{51}$$

while the angle of optical indicatrix rotation is the same as the tracing angle:

$$\tan 2\mathbf{z}_Z = \frac{2E_1E_2}{E_1^2 - E_2^2} = \tan 2\mathbf{j}, \quad \mathbf{z}_Z = \mathbf{j}. \tag{52}$$

In the other cases we will deal with multiplication of the polarization singularities.

### 5. CONCLUSIONS

We have proved for the first time that double-charged optical vortices can be generated using the electrooptic Kerr effect in single crystals as well as isotropic material media. We have analyzed all of the centrosymmetric point symmetry groups, including the Curie ones, with regard of possible generation of the optical vortices. In addition we have analyzed a number of non centrosymmetric groups of symmetry in which the Kerr effect is not masked by the Pockels one. It has been found that the principal condition for achieving this aim is optical uniaxiality of crystals and textures, or their optical isotropy. The second condition is a presence of inversion center among the symmetry elements or, otherwise, absence of any influence of the Pockels effect for the experimental geometries under interest. Still the most principal criterion for the generation of the doughnut mode is fulfillment of the condition $R_{44} = R_{11} - R_{12}$ imposed on the tensor Kerr coefficients for optically isotropic media and the condition $R_{66} = R_{11} - R_{12}$ for the uniaxial crystal. Otherwise the mixed screw-edge dislocation of the wave front will be generated and the optical vortex ring will become fractionalized.

We have found that the crystals, textures and isotropic media suitable for the generation of double-charged optical vortices should belong to the symmetry groups 622, 6mm, 6/mmm, 6, 6/m, $\infty/m$, $\infty$, $\infty 2$, $\infty mm$, $\infty/mmm$, $\infty/\infty/mmm$ and $\infty/\infty 2$. Hence, among the crystalline media, only hexagonal single crystals can be used for generating of the double-charged vortices. It is interesting that, according to our results, electrically induced optical vortices can be produced even in gases or liquids. Moreover, we have demonstrated that a transformation of the double-charged vortex into the single-charged one can be observed in the particular case when a longitudinal electric field is active owing to some nonzero Kerr coefficients. This process is accompanied by the appearance of a number of polarization singularities with the strength of topological defect equal to $\pm 1/2$, as well as a number of optical vortices with integer charges (namely, those equal to $\pm 1$).